\documentclass[fleqn,11pt,twoside]{article}
\usepackage{fullpage}
\usepackage{array}
\usepackage{listings}
\usepackage{graphicx}

\newtheorem{theorem}{Theorem}

\begin{document}
\lstset{												% general command to set parameter(s)
  basicstyle=\small, 						% print whole listing small
	keywordstyle=\color{black}\bfseries\underbar,			% underlined bold black keywords
	identifierstyle=, 						% nothing happens
	commentstyle=\color{white}, 	% white comments
	stringstyle=\ttfamily, 				% typewriter type for strings
	escapeinside={(*@}{@*)},
	stepnumber=1,
  numbersep=5pt,
	showstringspaces=false} 			% no special string spaces

\title{Implementation of float-float operators on graphics hardware}
\author{
  Guillaume Da Gra{\c c}a,  David Defour \\
  Dali, LP2A, Universit\'e de Perpignan, \\
  52 Avenue Paul Alduy,\\
	66860 Perpignan Cedex, France}

\maketitle      

\begin{abstract}
The Graphic Processing Unit (GPU) has evolved into a powerful and flexible processor. The latest graphic processors provide fully programmable vertex and pixel processing units that support vector operations up to single floating-point precision. This computational power is now being used for general-purpose computations. However, some applications require higher precision than single precision. This paper describes the emulation of a 44-bit floating-point number format and its corresponding operations. An implementation is presented along with performance and accuracy results.
\end{abstract}

%\begin{keywors}
%GPGPU, IEEE-754, floating-point arithmetic, accuracy.
%\end{keywords}

\section{Introduction}
There is significant interest in using graphics processing units (GPUs) for general purpose programming.  These GPUs have an explicitly parallel programming model and deliver much higher performance for some floating-point workloads when compared to CPUs. This explains the growing concern in using a graphic processor as a stream processor for executing highly parallel applications. 

\subsection{The graphics pipeline}
\begin{figure}[!htb]
	\begin{center}
	 \includegraphics*[width=0.9\textwidth]{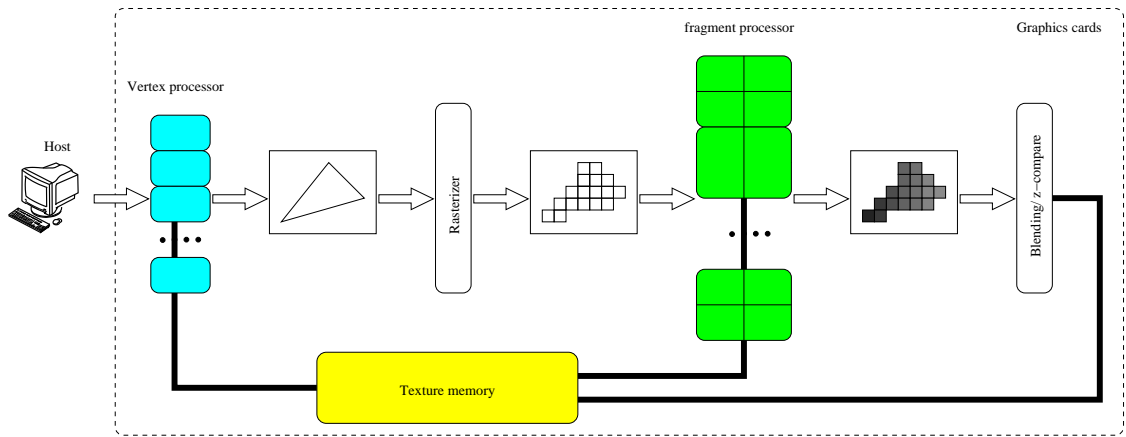}
	\caption{This figure illustrates the current graphics pipeline found in recent PC graphics cards. \label{gp}}
	\end{center}
\end{figure}

Data processed by the GPU are mainly pixel, geometric objects and elements that create the final picture in the frame buffer. These objects require an intensive computation before getting the final image. This computation is done within the "Graphics Hardware Pipeline". The pipeline contains several steps in which the 3D application sends a sequence of vertices to the GPU that are batched into geometric primitives (polygons, lines, points).  These vertices are processed by the programmable vertex processor that has the ability to perform mathematical operations. Then the resulting primitives are sent to the programmable fragment processor. Fragment processors require the same mathematical operation as the vertex processors, plus some texturing operations. A representation of this pipeline is shown in figure \ref{gp}.

The computational workhorse of the GPU is located within the 2 programmable processors: vertex processors and fragment processors. The amount of these processors embedded in GPUs has greatly increased over the years; for example the latest Nvidia 7800GTX chip integrates 8 vertex shaders and 24 pixel shaders. In this chip, each vertex shader is made up of 1 multiply and accumulate (MAD) unit and 1 special function unit that can compute log, exp, sin, cos. The implementation of a similar unit is detailed in \cite{Obe05}. Each pixel shader of the 7800GTX consists of 2 consecutive MADs, therefore the 7800GTX is able to execute 112 floating-point operations in single precision per clock cycle at a peak rate.

\subsection{Representation formats available in GPUs}
Vertex and pixel shaders were originally composed of fixed-point operators that have evolved into partial support of the IEEE-754 single precision floating-point format. For example, the Nvidia GeForce 6 series offers a 32-bit format similar to single precision. The other main GPU manufacturer, ATI, integrated a 32-bit floating-point arithmetic required by shader version 3.0 in their latest chips, the X1k series. Older ATI hardware performed floating-point operations on a 24-bit format in spite of the fact that they stored values in the IEEE standard 32-bit format as described in table \ref{fpformat}

\begin{table}[h]
\begin{center}
\begin{tabular}{|l|c|c|c|c|}
\hline
Format name	&Sign&	Exponent&	Mantissa&	Support for special \\
& & & & values (NaN, Inf, …)\\ \hline \hline
Nvidia 16-bit&	1&	5&	10&	Yes\\ \hline
Nvidia 32-bit&	1&	8&	23&	Yes\\ \hline
ATI 16-bit&	1&	5&	10&	No\\ \hline
ATI 24-bit&	1&	7&	16&	No\\ \hline
ATI 32-bit&	1&	8&	23&	? / \emph{not tested} \\ \hline
\end{tabular}
\caption{Floating-point format currently supported by the Nvidia and the ATI.\label{fpformat}}
\end{center}
\end{table}

In addition to these formats, current GPUs support other data types that are of lower precision. Therefore, applications where accuracy is paramount are not well suited for a GPU execution due to the lack of the double precision format, the non uniformity within the floating-point format and the non-respect of the IEEE-754 requirement (such as rounding or denormal number which are typically flushed to zero \cite{Cebenoyan2005}). 

The purpose of this article is to propose a software solution to the limitation of precision in floating-point operations and storage. This solution consists of an implementation of a \emph{float-float} format which doubles the hardware accuracy. This corresponds to a 44-bit precision on Nvidia architecture. 

\subsection{Outline of this paper}
Based on the above clarification, hence section 2 presents similar work related to multiprecision operators. Section 3 describes the floating-point arithmetic available in current GPU. Section 4 proposes a representation and the algorithms for the basic multiprecision operations. Section 5 describes our initial implementation used to conduct tests and comparisons which results are discussed in section 6.

\section{Related work}
Many modern processors obey the IEEE-754 \cite{ieee754} standard for floating-point arithmetic, which define the single and double precision format. For some applications, however, the precision provided by the hardware operators does not suffice. These applications include large scale simulation, number theory and multi-pass algorithm like shading, lighting \cite{Strzodka2002}.

Applications encountering accuracy problems are commonly executed on a CPU. As a consequences, most of the research done to develop a \emph{multiprecision} format (a representation format with a precision higher than the one available in the hardware) were done on the CPU. 

\subsection{CPU related work}
Many software libraries were proposed to address the precision issue in hardware limitation. These libraries emulate arithmetic operators with higher precision than the one provided by the hardware. They either use integer units or floating-point units, depending on the internal representation of their number.

\subsubsection*{Libraries based on an integer representation}
All the libraries in this category, internally represent multiprecision numbers as an array of integers, which are machine numbers (usually 32-bit or 64-bit) to store the significant of the multiprecision numbers. It is the case for GMP \cite{GMPweb}, on top of which several other libraries are built (see MPFR\cite{MPFRweb}). These libraries allow the user to dynamically set the precision of the operation during the execution of the program. However some other libraries \cite{Ziv2002,russie02}  set the precision at compilation time to get higher performance. 

\subsubsection*{Libraries based on a floating-point representation}
The actual trend of CPUs is to have highly optimized floating-point operators. Some libraries, such as the MPFUN\cite{Bailey95}, exploit these floating-point operators by using an array of floating-point number. 

Others libraries represent multiprecision numbers as the unevaluated sum of several double-precision FP numbers such as Briggs' double-double \cite{Bri1998}, Bailey's quad-doubles \cite{Bailey2001} and Daumas' floating-point expansions \cite{Dau99}. This representation format is based on the IEEE-754 features that lead to simple algorithms for arithmetic operators. However this format is confined to low precision (2 to 3 floating-point number) as the complexity of algorithms increases quadratically with the precision.

\subsection{GPU related work}
The available precision provided through the GPU's graphical pipeline is limited; for example the color channel is usually represented with a 8-bit number. Before the introduction of the shader 3.0 that requires support of 32-bit floating-point numbers, developers and researchers that were facing accuracy problems developed software solutions to extend the hardware precision.

For example, Strzodka \cite{Strzodka2002} proposed a 16-bit fixed-point representation and operation out of the 8-bit fixed-point format. In his work, two 8-bit numbers were used to emulate 16-bit. The author claimed that operators in his representation format were only $50\%$ slower than normal operators, however no measured timings were provided. Strzodka recently implemented a FEM algorithm in double precision on GPU\cite{Goeddeke2005}, nethertheless double precision computation were sent to the CPU. This method involes time consuming memory transfert.

\section{Floating-point arithmetic on GPUs}
\label{fpgpu}
Floating-point computations on GPUs are often called into question. Current GPUs do not strictly conform to the IEEE-754 floating-point standard. This produces differences between the same computation performed on the GPU and the CPU, and among GPUs themselves. Floating-point computation details vary with GPU models and they are kept secret by GPU manufacturers. 

Recently, one tool has been developed to understand some of the details of the floating-point arithmetic for a given GPU \cite{Hillesland2004}. We executed this tool on GPUs, which is an adaptation of Paranoia, and got some resulting errors reported in table \ref{paranoia}.

\begin{table}[h]
\begin{center}
\begin{tabular}{|l|c|c|c|c|}
\hline
Operation&			Exact &	Chopped&	R300&	NV35 \\ 
&			rounding &	&	&	 \\ \hline \hline
Addition&				[-0.5, 0.5]&	(-1, 0]&	[-1.0, 0.0]&	[-1.0, 0.0]\\ \hline
Substraction&		[-0.5, 0.5]&  (-1, 1)&	[-1.0, 1.0]&	[-0.75, 0.75]\\ \hline
Multiplication&	[-0.5, 0.5]&	(-1, 0]&	[-0.989, 0.125]&	[-0.782, 0.625]\\ \hline
Division&				[-0.5, 0.5]&	(-1, 0]&	[-2.869, 0.094]&	[-1.199, 1.375]\\ \hline
\end{tabular}
\caption{Floating-point error from the execution of the paranoia Test \cite{Hillesland2004} \label{paranoia}}
\end{center}
\end{table}

Table \ref{paranoia} shows us that the addition is truncated after the last bit on both ATI R300 and Nvidia NV35. The subtraction benefits from a guard bit on Nvidia processors and not on ATI. This property is very important for numerical algorithm as we will see later on in this paper. The multiplication is faithfully rounded on both the ATI and the Nvidia. Because GPUs do not provide a division instruction, every division is performed as a reciprocal followed by a multiplication; thereby the floating-point error for the division incurs double floating-point errors. 

\section{Proposed format}
\label{proposedformat}
The proposed format is an adaptation of the double-double format described in \cite{Bri1998}. For our algorithm we chose to represent multiprecision numbers as the unevaluated sum of 2 floating-point numbers handled in hardware. The type of hardware representation used is described in table \ref{fpformat}.  

In the core of the GPU, the graphical pipeline, is made up of several computational units. These processing units are not design to efficiently perform tests and comparisons, therefore whenever it is possible, we should avoid tests even at the expense of extra computations. In addition, software that does not use branches remains compatible with older GPU. In our case, two versions of Add12 algorithms exist \cite{She97}; one with one test and another one, that should be preferred, with 3 extra floating-point operations. 

\subsection{Mathematical background}
In this section we present some basic properties and algorithms of the IEEE floating-point arithmetic used in our format. In this paper we assume that GPUs have a guard bit for the addition/subtraction with a faithful rounding as it seems to be the case with latest Nvidia chips. The multiplication will behave as observed in section \ref{fpgpu}. This assumption conforms to the actual trends follow by the GPU and the shader model 3.0. For any mathematical operator $+, -, *, /$, we use $\oplus, \ominus, \otimes, \oslash$ to represent the hardware operator that may involve a rounding error.

\begin{theorem}[Sterbenz lemma (\cite{Gol91} Th. 11)]
If subtraction is performed with a guard digit, and $y/2 \leq x \leq 2y$, then $x \ominus y$ is computed exactly.
\end{theorem}

\begin{theorem}[Add12 theorem (Knuth \cite{Knu73})]
Let $a$ and $b$ be normalized floating-point numbers. The following algorithm computes $s=a \oplus b$ and $r= (a+b)- s$ such that $s + r = a+b$ exactly, provided that no exponent overflow or underflow occurs. 
\begin{lstlisting}[frame=single]
Add12(a, b)
s = a (*@$\oplus$@*) b
v = s (*@$\ominus$@*) a
r = (a(*@$\ominus$@*)(s(*@$\ominus$@*)v))(*@$\oplus$@*)(b(*@$\ominus$@*)v)
return (s,r)
\end{lstlisting}

\end{theorem}

\paragraph{proof}
The proof of correctness of the Add12 algorithm, in an environment with a correctly rounded arithmetic, is described in Knuth \cite{Knu73} or Shewchuk \cite{She97} articles. However, by examining these proofs, one can observe that they are based on Sterbenz lemma. This lemma only requires a guard bit to be true, therefore the Add12 theorem is true on Nvidia hardware.

\begin{theorem}[Split theorem (Dekker \cite{Dek71})]
Let $a$  be $p$-bit floating-point number, where $p\geq3$. Choose a splitting point $s$ such that $p/2 \leq s \leq p-1$. Then the following algorithm will produce a $(p-s)$-bit value $a_{hi}$ and a non-overlapping $(s)$-bit value $a_{lo}$ such that $|a_{hi}| \geq |a_{lo}|$ and $a=a_{hi} + a_{lo}$.
\begin{lstlisting}[frame=single]
SPLIT( a )
1 c = ((*@$2^s$@*) (*@$\oplus$@*) 1) (*@$\otimes$@*) a
2 (*@$a_{big}$@*) = c (*@$\ominus$@*) a
3 (*@$a_{hi}$@*) = c (*@$\ominus$@*) (*@$a_{big}$@*)
4 (*@$a_{lo}$@*) = a (*@$\ominus$@*) (*@$a_{hi}$@*)
5 return ((*@$a_{hi} ~,~ a_{lo}$@*))
\end{lstlisting}
\end{theorem}
 
\paragraph{Proof}
This proof is an adaptation of the proof from \cite{She97} to fit the condition observed on GPUs. 
Line 1 is equivalent to computing $2^s a \oplus a$ , because multiplying by a power of two only changes its exponent. The addition is subject to rounding, so we have $c=2^s a + a + err(2^s a \oplus a)$.
Line 2 is subject to rounding, so $a_{big} = 2^s a + err(2^s a \oplus a) + err(c \ominus a)$. Both $|err(2^s a \oplus a)|$ and $|err(c \ominus a)|$ are bounded by $ulp(c)$, so the exponent of $a_{big}$ can only be larger than that of $2^s a$ if every bit of the significand of $a$ is nonzero except the last two bits. By manually checking the behavior of SPLIT in these 4 cases, one can verify that the exponent of $a_{big}$ is never larger than that of $2^s a$. Then $|err(c \ominus a) | \leq ulp(2^s a)$, and so the error term $err(c \ominus a)$ is expressible in $s$ bits.

By Sterbenz lemma line 3 and 4 are calculated exactly. It follows that $a_{hi}=a-err(c \ominus a)$ and $a_{lo} = err (c \ominus a)$; the latter is expressible in $s$ bits. Either $a_{hi}$ has the same exponent as $a$ either $a_{hi}$ has an exponent one greater than that of $a$ and in both case $a_{hi}$ is expressible in $p-s$ bits.

\begin{theorem}[Mul12 theorem (Dekker \cite{Dek71})]
Let $a$ and $b$ be $p$-bit floating-point numbers, where $p \geq 6$. The following algorithm produces two floating point numbers $x$ and $y$ as results such that $a \cdot b = x + y$, where $x$ is an approximation to $a \cdot b$ and $y$ represents the roundoff error in the calculation of $x$.
\begin{lstlisting}[frame=single]
Mul12(a,b)
1 x = a (*@$\otimes$@*) b
2 ((*@$a_{hi}~,~ a_{lo}$@*)) = SPLIT (a)
3 ((*@$b_{hi}~,~ b_{lo}$@*)) = SPLIT (b)
4 err1 = x (*@$\ominus$@*) ((*@$a_{hi}\otimes b_{hi}$@*))
5 err2 = err1 (*@$\ominus$@*) ((*@$a_{lo}\otimes b_{hi}$@*))
6 err3 = err2 (*@$\ominus$@*) ((*@$a_{hi}\otimes b_{lo}$@*))
7 y = ((*@$a_{lo} \otimes b_{lo} ) \ominus$@*) err3
8 return (x,y)
\end{lstlisting}
\end{theorem}

\paragraph{Proof}
Line 1 computes $x=ab + err(a \otimes b)$ with $err(a \otimes b)$ the rounding error of the multiplication. One can noticed that all the other multiplications and subtractions are exact and compute $y=-err(a \otimes b)$.
 
\begin{theorem}[Add22 theorem]
\label{add22algo}
Let be $ah+al$ and $bh+bl$ the float-float arguments of the following algorithm:
\begin{lstlisting}[frame=single]
Add22(ah, al, bh, bl)
1 r = ah (*@$\oplus$@*) bh
2 if |ah| (*@$ \geq $@*) |bh| then 
3    s = (((ah (*@$\ominus$@*) r) (*@$\oplus$@*) bh) (*@$\oplus$@*) bl) (*@$\oplus$@*) al
4 else
5    s = (((bh (*@$\ominus$@*) r) (*@$\oplus$@*) ah) (*@$\oplus$@*) al) (*@$\oplus$@*) bl
6 (rh, rl) = add12(r, s)
7 return (rh, rl)
\end{lstlisting}
The two floating-point numbers $rh$ and $rl$ returned by the algorithm verifies
$$
rh + rl =(ah+al)+(bh+bl) + \delta
$$
Where $\delta$ is bounded as follows:
$$\delta \leq max(2^{-24} \cdot |al+bl| , 2^{-44} \cdot |ah + al + bh + bl|)$$
\end{theorem}

\begin{theorem}[Mul22 theorem]
Let be $ah +bl$ and $bh+bl$ the float-float arguments of the following algorithm:
\begin{lstlisting}[frame=single]
Mul22(ah, al, bh, bl)
1 (t1, t2) = Mul12(ah, bh)
2 t3 = ((ah (*@$\otimes$@*) bl) (*@$\oplus$@*) (al (*@$\otimes$@*) bh)) (*@$\oplus$@*) t2
3 (rh, rl) = Add12(t1,t2)
4 return (rh, rl)
\end{lstlisting}
The result $rh + rl$ returned by the algorithm verifies
$$rh + rl=((ah+al)*(bh+bl))*(1+\epsilon)$$
Where $\epsilon$ is bounded as follows:
$$|\epsilon| \leq 2^{-44}$$
\end{theorem}

\paragraph{Proof:}
The detailed proof of the Add22 and the Mul22 theorem were proposed by Lauter in \cite{Lauter2005} for the particular cases of double-double format on an IEEE compliant architecture. The proof of these 2 theorems with GPU conditions (single precision, faithful rounding and guard bit) is very similar and is therefore not detailed here for the sake of clarity.

\section{Implementation}
We developed a Brook \cite{Buck2004} implementation of the float-float format and of Add12, Add22, Split, Mul12, Mul22 algorithms. Brook is a high level programming language designed for general purpose programming on GPUs. This language allows us to test our algorithms with ease over various systems, drivers and graphics hardware with minor modifications.

During our implementation, we observed that the DirectX version generated by Brook were performing forbidden floating-point optimization. These floating-point optimizations were not noticed with the OpenGL version. For example, the sequence of operations that compute the rounding error $r = ((a \oplus b) \ominus a)$ was replaced by $r = b$. To overcome this problem, we had to apply hand correction on the fragment program generated by Brook.

%As graphic hardware's are dealing with pixel information and not with single floating-point values, we developed several versions for each algorithm. The different tested versions were implemented the possibilities for storing input and output stream. Figure XXX represent these versions. 

%However we will note that GPU elements of a stream might be processes out of order. This property of GPU does not allow us to represent the float-float format within a 1 dimensional array. In addition, in a CPU, data are represented in an array whereas in a GPU, data are represented in a matrix. 

%We tested the different implementation of the algorithms on matrix on the GPU and compared it with an array with the same amount of data on the CPU.

\section{Results and performance}
It is quite difficult to compare the performances of GPU and CPU operators. CPUs already have data stored in the memory hierarchy whereas GPUs have to download data from main memory to its local memory before processing it. To make fair comparisons, we compared float-float algorithms to basic single precision operations (addition, multiplication, multiply and add) on a CPU and on a GPU. For clarity we normalized results to the time of 4096 additions. For each version we tested different sizes of data set. Tests were done on a Nvidia 7800GTX graphics card with 256 MB and on a Pentium IV HT 3.2 Ghz.

\begin{table}[h]
\begin{center}
\begin{tabular}{|l|c|c|c|c|c|c|c|}
\hline
Size &	Add	&Mull	&Mad	&Add12	&Mul12	&Add22	&Mul22 \\ \hline\hline
4096	&1,00	&0,97	&1,00	&1,09	&1,57	&1,55	&1,54\\ \hline
16384	&1,11	&1,11	&1,15	&1,20	&1,87	&1,73	&2,02\\ \hline
65536	&1,55	&1,58	&1,69	&1,64	&2,09	&2,87	&2,94\\ \hline
262144	&3,55	&3,40	&3,44	&3,74	&3,99	&7,15	&7,47\\ \hline
1048576	&10,64	&10,74	&10,75	&10,79	&14,64	&23,92	&24,64\\ \hline
\end{tabular}
\caption{Timing comparison of float-float operators executed on the GPU. The time is normalized on the single addition of 4096 data. \label{tbl:time_gpu}}
\end{center}
\end{table}

\begin{table}[h]
\begin{center}
\begin{tabular}{|l|c|c|c|c|c|c|c|}
\hline
Size &	Add	&Mull	&Mad	&Add12	&Mul12	&Add22	&Mul22\\ \hline\hline
4096	&1,00	&0,98	&1,35	&1,52	&2,86	&11,71	&4,12\\ \hline
16384	&3,88	&3,88	&3,46	&6,04	&17,86	&47,93	&17,62\\ \hline
65536	&17,13	&16,20	&17,67	&28,35	&49,14	&192,10	&69,33\\ \hline
262144	&68,77	&66,68	&77,10	&100,10	&187,49	&760,65	&272,13\\ \hline
1048576	&269,49	&267,88	&312,45	&419,84	&1027,62	&3083,74	&1091,59\\ \hline
\end{tabular}
\caption{Timing comparison of float-float operators executed on the CPU. The time is normalized on the single addition of 4096 data. \label{tbl:time_gpu}}
\end{center}
\end{table}

We have not reported the difference of execution time between CPUs and GPUs because such comparison is meaningless. However to give an idea, we measured that sending data to the GPU, executing the 4096 additions and getting back the results on the CPU correspond to 100 times the execution time of the same 4096 addition on the CPU. This overhead mainly come from the use of the bus of the system to send and to get back data. Therefore GPU will faster than CPU if many operations will be done on the same large set of data.  

The difference of time between small and large data set is higher for the CPU than for the GPU. This difference is of 25 for GPU and 3000 for CPU. This means that GPUs are more efficient at performing the same operation over a large set of data. The Add22 times on CPU is much higher than other operations. An interpretation could be that the test in the Add22 algorithm \ref{add22algo} is time consuming compared to normal operations as it breaks the execution pipeline. 

We observe that the execution time of the addition, the multiplication, the multiply and accumulate and the Add12 have the same cost on GPU. The algorithm Add22 and Mull22 cost twice as much as basic operations. This proves that GPU drivers are very efficient at merging operations of different execution loops. This also signifies that the cost of these operations could be higher when used in a real program.

\subsection{Accuracy}
\label{sec:accuracy}
We ran our algorithms on $2^{24}$ randomly generated test vectors and we collected the maximum observed error with the help of MPFR\cite{MPFRweb}. For these tests, we excluded denormal input numbers and special cases numbers as there are not fully supported by the targeted hardware.

\begin{table}[h]
\begin{center}
\begin{tabular}{|l|c|}
\hline
Operation&			Error max 	 \\ \hline \hline
Add12&	-48.0		 \\ \hline
Mul12&		(exact) \\ \hline
Add22&  -33.7  \\ \hline
Mul22&	-45.0  \\ \hline
\end{tabular}
\caption{Measured accuracy on our GPU float-float implementation \label{tbl:accuracy}}
\end{center}
\end{table}

The first observation we can make is that the reported accuracy is different from the theoretical one. We proved in section \ref{proposedformat} that if GPUs have a guard bit then Add12 will be exact. Our initial tests show us that GPUs behave as if they have a guard. However in a very special case the error is higher than expected. This happens when two floating point numbers of opposite signs are summed up together and when their mantissa are not overlapping in a certain way. Further investigations have to be done to locate and correct the problem. This problem is also the cause of the bad accuracy result of the Add22 algorithm.

\section{Conclusion and future work}
In our work, we have described a general framework for the implementation of software emulation of floating-point numbers with 44 bits of accuracy. The implementation is based on Brook and allows simple and efficient addition, multiplication and storage of floating-point number. The representation range of this format is similar to single precision. These high precision operations naturally require more texture memory and computing time. However, they proved to remain fast enough to be used in precise sensitive parts of real-time multipass algorithms.

During our test, we noticed that Brook was well suited for fast prototyping of functions; however, as with every high level languages, we were unable to have fine control over GPUs instructions. In particular, it was not possible to control how and when data were stored, transferred and used within the GPU. Therefore, we are currently working on an OpenGL and Cg version of these functions. We hope that it will lead to an improvement in performance. We are also investigating to set a solution to the accuracy problem described in section \ref{sec:accuracy}. Using float-float representation number in compensated algorithms has been shown to be more efficient in term of performance for comparable accuracy. Adapting compensated algorithm to GPU is part of our future investigation.

\bibliographystyle{plain}
\bibliography{my-biblio,biblio-general}
\end{document}